%% file: ntx_date2019.tex
\documentclass[conference, hidelinks]{IEEEtran}
\IEEEoverridecommandlockouts
\usepackage[utf8]{inputenc}
\usepackage[T1]{fontenc}
\usepackage[english]{babel}

\usepackage{cite}
\usepackage{amsmath,amssymb,amsfonts}
\usepackage{algorithmic}
\usepackage{graphicx}
\usepackage{textcomp}
\usepackage{xcolor}
\usepackage{booktabs}
\usepackage{tabularx}
\usepackage[para]{threeparttable}
\usepackage{multirow}
\usepackage{multicol}
\usepackage[binary-units]{siunitx}
\usepackage[acronym]{glossaries}
\usepackage{hyperref}

\newcolumntype{C}{>{\centering\arraybackslash}X}
\newcolumntype{R}{>{\raggedleft\arraybackslash}X}
\newcolumntype{x}[1]{>{\centering\arraybackslash\hspace{0pt}}p{#1}}

\newcommand{\ubold}{\fontseries{b}\selectfont}
\robustify\ubold
\newcolumntype{k}[1]{S[
  table-format=#1,
  detect-weight,
  input-symbols={()},
]}

\newcommand{\revb}[1]{#1}

\newcommand{\figref}[1]{{Figure~\ref{#1}}}
\newcommand{\tabref}[1]{{Table~\ref{#1}}}

\DeclareSIUnit\op{op}

\DeclareSIUnit\OP{op}
\DeclareSIUnit\OPs{\OP\per\s}
\DeclareSIUnit\OPsW{\OP\per\s\per\watt}
\DeclareSIUnit\GOPs{\giga\OPs}
\DeclareSIUnit\GOPsW{\giga\OPsW}
\DeclareSIUnit\TOPs{\tera\OPs}
\DeclareSIUnit\TOPsW{\tera\OPsW}

\DeclareSIUnit\FLOP{flop}
\DeclareSIUnit\FLOPs{\FLOP\per\s}
\DeclareSIUnit\FLOPsW{\FLOP\per\s\per\watt}
\DeclareSIUnit\GFLOPs{\giga\FLOPs}
\DeclareSIUnit\GFLOPsW{\giga\FLOPsW}
\DeclareSIUnit\TFLOPs{\tera\FLOPs}
\DeclareSIUnit\TFLOPsW{\tera\FLOPsW}

\graphicspath{{figures/}}

\makeglossaries
\input{acronyms}

\begin{document}

%%%%%%%%%%%%%%%
%%   TITLE   %%
%%%%%%%%%%%%%%%

\title{
    NTX: An Energy-efficient Streaming Accelerator for Floating-point Generalized Reduction Workloads in 22\,nm FD-SOI
    \thanks{This work has been supported by Microsoft Research under the project ``Enabling Practical, Efficient and Large-Scale Computation Near Data to Improve the Performance and Efficiency of Data Center and Consumer Systems'' with MRL contract number 2017-044.}
    \thanks{This work has received funding from the European Union’s Horizon 2020 research and innovation programme under grant agreement No 732631, project ``OPRECOMP''.}
}

\author{
    \IEEEauthorblockN{Fabian Schuiki}
    \IEEEauthorblockA{
        \textit{IIS, ETH Zürich}\\
        Zürich, Switzerland \\
        fschuiki@iis.ee.ethz.ch
    }
    \and
    \IEEEauthorblockN{Michael Schaffner}
    \IEEEauthorblockA{
        \textit{IIS, ETH Zürich}\\
        Zürich, Switzerland \\
        schaffner@iis.ee.ethz.ch
    }
    \and
    \IEEEauthorblockN{Luca Benini}
    \IEEEauthorblockA{
        \textit{IIS, ETH Zürich}\\
        Zürich, Switzerland \\
        \textit{DEI, University of Bologna}\\
        Bologna, Italy \\
        lbenini@iis.ee.ethz.ch
    }
}

\maketitle

%%%%%%%%%%%%%%%%%%
%%   ABSTRACT   %%
%%%%%%%%%%%%%%%%%%

\begin{abstract}
Specialized coprocessors for Multiply-Accumulate (MAC) intensive workloads such as Deep Learning are becoming widespread in SoC platforms, from GPUs to mobile SoCs. In this paper we revisit NTX (an efficient accelerator developed for training Deep Neural Networks at scale) as a generalized MAC and reduction streaming engine. The architecture consists of a set of 32\,bit floating-point streaming co-processors that are loosely coupled to a RISC-V core in charge of orchestrating data movement and computation. Post-layout results of a recent silicon implementation in 22\,nm FD-SOI technology show the accelerator's capability to deliver up to 20\,Gflop/s at 1.25\,GHz and 168\,mW. Based on these results we show that a version of NTX scaled down to 14\,nm can achieve a 3$\times$ energy efficiency improvement over contemporary GPUs at 10.4$\times$ less silicon area, and a compute performance of 1.4\,Tflop/s for training large state-of-the-art networks with full floating-point precision. An extended evaluation of MAC-intensive kernels shows that NTX can consistently achieve up to 87\% of its peak performance across general reduction workloads beyond machine learning. Its modular architecture enables deployment at different scales ranging from high-performance GPU-class to low-power embedded scenarios.

\end{abstract}


%%%%%%%%%%%%%%%%%%
%%   SECTIONS   %%
%%%%%%%%%%%%%%%%%%

\input{sec_intro}
\input{sec_arch}
\input{sec_results}
\input{sec_relwork}

\input{sec_conc}

\bibliographystyle{IEEEtran}
\bibliography{IEEEabrv,ref}

\end{document}

%% file: acronyms.tex
\newacronym{agu}{AGU}{Address Generation Unit}
\newacronym{asmc}{ASMC}{Application Specific Memory Cube}
\newacronym{cnn}{CNN}{Convolutional Neural Network}
\newacronym{dag}{DAG}{Directed Acyclic Graph}
\newacronym{dnn}{DNN}{Deep Neural Network}
\newacronym{fdsoi}{FD-SOI}{Fully Depleted Silicon On Insulator}
\newacronym{flops}{flops}{Floating-Point Operations}
\newacronym{fmac}{FMAC}{Floating-Point Multiply and Accumulate}
\newacronym{fma}{FMA}{Fused Multiply and Add}
\newacronym{fpu}{FPU}{floating-point unit}
\newacronym{fp}{FP}{floating-point}
\newacronym{gru}{GRU}{Gated Recurrent Unit}
\newacronym{hbm}{HBM}{High Bandwidth Memory}
\newacronym{hmc}{HMC}{Hybrid Memory Cube}
\newacronym{hpc}{HPC}{High Performance Computing}
\newacronym{hwl}{HWL}{Hardware Loop}
\newacronym{innovus}{Innovus}{Cadence Innovus}
\newacronym{lim}{LiM}{Logic in Memory}
\newacronym{lob}{LoB}{Logic Base}
\newacronym{lstm}{LSTM}{Long Short-Term Memory}
\newacronym{mac}{MAC}{Multiply and Accumulate}
\newacronym{mlp}{MLP}{Multi-Layer Perceptron}
\newacronym{mmu}{MMU}{Memory Management Unit}
\newacronym{pcs}{PCS}{Partial Carry-Save}
\newacronym{pim}{PiM}{Processor-in-Memory}
\newacronym{pue}{PUE}{Power Usage Effectiveness}
\newacronym{relu}{ReLU}{Rectified Linear Unit}
\newacronym{rmse}{RMSE}{Root Mean Squared Error}
\newacronym{rnn}{RNN}{Recurrent Neural Network}
\newacronym{rtl}{RTL}{Register Transfer Level}
\newacronym{sgd}{SGD}{Stochastic Gradient Descent}
\newacronym{tcdm}{TCDM}{Tightly-Coupled Data Memory}
\newacronym{tlb}{TLB}{Translation Look-aside Buffer}
\newacronym{vfs}{VFS}{Voltage and Frequency Scaling}

%% file: sec_intro.tex
% ==============================================================================
\section{Introduction}
\label{sec:intro}

Specialized accelerators for parallel MAC intensive workloads are becoming essential platforms ranging from mobile SoCs to high-performance GPUs, due to the widespread diffusion of \glspl{dnn} into various classification and recognition tasks \cite{Goodfellow2016, Sze2017}. Yet most such accelerators are narrowly specialized for inference only \cite{Sze2017,Gao2017, azarkhish2017neurostream, Du2015}. Acceleration of training and more general MAC-intensive workloads has only received moderate attention thus far \cite{venkataramani2017scaledeep, luo2017dadiannao, kim2016neurocube} and is still mainly carried out using GPUs \cite{nvidia2017volta}.

\begin{figure}
  \centering
  \includegraphics[width=1\linewidth]{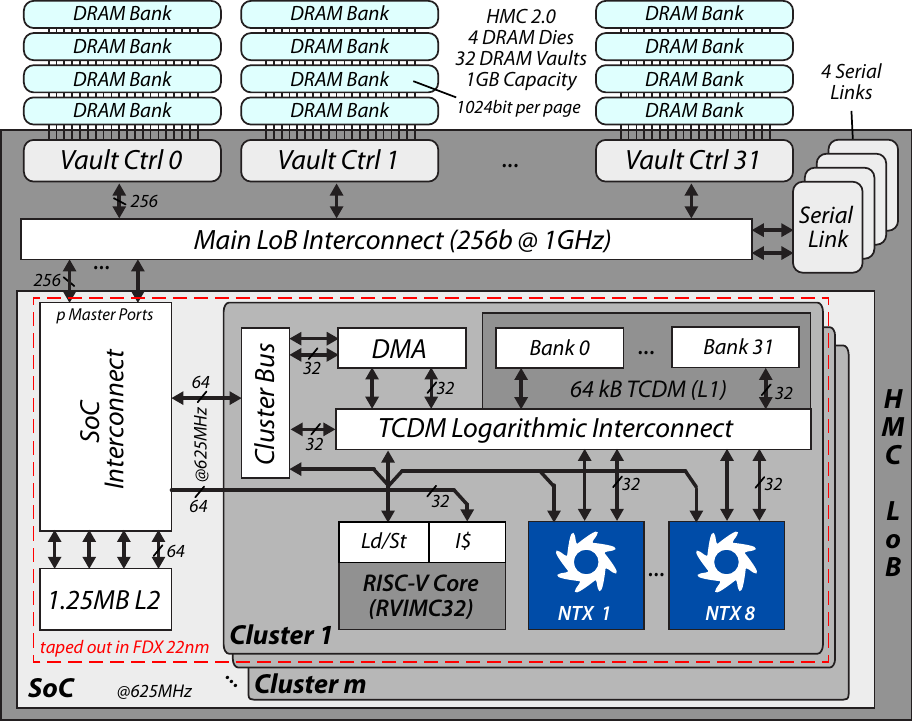}
  \caption{Top-level block diagram of one \gls{hmc} enhanced with $m$ processing clusters. The \gls{lob} contains the vault controllers, main interconnect, and the four serial links that lead off-cube. The proposed processing clusters attach directly to the main interconnect and gain full access to the \gls{hmc}'s memory space and the serial links. Each cluster consists of a DMA unit, a \gls{tcdm}, and one or more RISC-V processor cores augmented with NTX streaming co-processors.}
  \label{fig:arch:system}
\end{figure}

\begin{figure*}[t]
  \centering
  \includegraphics[width=1\linewidth]{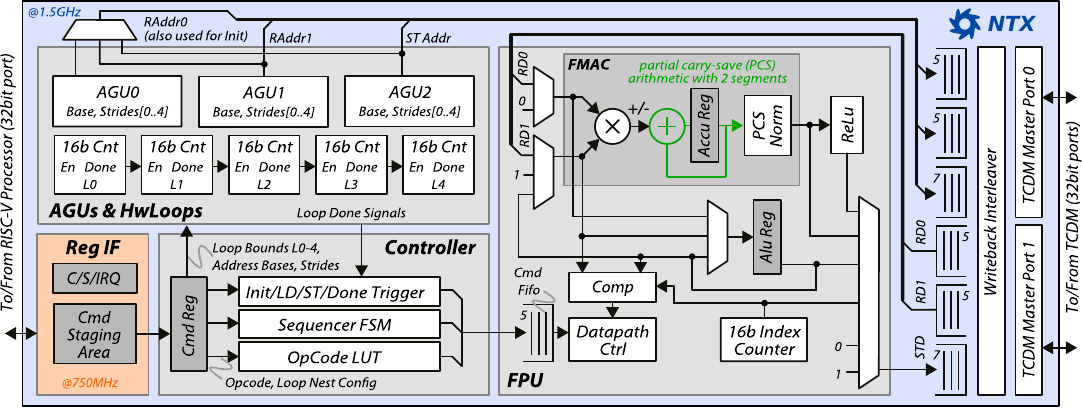}
  \caption{Block diagram of the NTX accelerator. It contains 5 hardware loops; 3 address generator units; a double-buffered command staging area in the register interface; a main controller; and a FPU with a comparator, index counter (for argmax calculations) and a fast \gls{fmac} unit. The employed depths for all FIFOs are indicated and have been determined in simulations for a TCDM read-latency of 1 cycle.}
  \label{fig:arch:nsl}
\end{figure*}

Training of \glspl{dnn} and general MAC-intensive workloads incur additional complexity due to additional data dependencies and higher accuracy requirements when compared to inference. At the same time the parameter memory footprint of typical \glspl{dnn} has grown rapidly from a few \si{\mega\byte} to several tens or hundreds of \si{\mega\byte} over the last years \cite{Szegedy2015, he2016deep}. The corresponding training data sets are bound to grow as well, since larger networks require more training data in order to reach good generalization performance. These observations suggest that \gls{pim} architectures that leverage lower access latencies and efficient data movement mechanisms should be closer investigated in the context of \gls{dnn} training.

In this work we revisit NTX, a recently proposed \gls{pim} accelerator for deep learning applications \cite{schuiki2018scalable}\footnote{This invited paper summarizes the \emph{IEEE Transactions on Computers} article \cite{schuiki2018scalable} (in press), and extends it with results from a recent tape out in 22\,nm and additional kernel benchmarks beyond the realm of deep learning.}. In particular we provide results of a recently taped out variant that consists of a single processing cluster with 1 RISC-V core and 8 NTX accelerators, capable of delivering \SI{20}{\GFLOPs} at \SI{1.25}{\giga\hertz} in 22\,nm FD-SOI technology. Based on post-layout simulation results, we show that a \gls{hmc} extended with NTX can achieve a compute performance of \SI{1.4}{\TFLOPs} at an efficiency of \SI{63.5}{\GFLOPsW} while training large state-of-the-art networks at full 32\,bit floating-point precision. In addition, we investigate NTX's applicability to and efficiency operating on more general kernels beyond machine learning, such as basic linear algebra subprograms and stencil codes. These are important methods in many domains, e.g. for solving least squares or finite difference problems in image and signal processing applications such as simultaneous localization and \cite{kerl2013}, optical flow estimation and inpainting \cite{krishnan2011, koutis2011}, and weather and seismic modeling \cite{gysi2015modesto, krueger2011hardware}. The contributions of this paper are:

\begin{itemize}
	\item post-layout timing and power results based on a design recently taped out in 22\,nm technology; and
	\item an extended performance analysis of general reduction kernels beyond deep learning.
\end{itemize}

We find that NTX is a competitive solution for general reduction applications. Its modular architecture enables deployment at scales other than training in a data center, for example as an accelerator for low power data analytics on edge devices.

%% file: sec_arch.tex
\section{Architecture}
\label{sec:arch}

In the following we provide a short overview of the NTX architecture outlined in more details in \cite{schuiki2018scalable}. The \glsfirst{lob} of a \gls{hmc} offers a unique opportunity to introduce a \gls{pim} as depicted in \figref{fig:arch:system}. The memory dies are subdivided into vertically connected vaults, with corresponding memory controllers on the \gls{lob} connected to the serial links via the main interconnect. Our architecture consists of multiple processing clusters attached to this interconnect which thus gain full access to the entire \gls{hmc} memory space, and sibling \glspl{hmc} attached via the serial links.

% ------------------------------------------------------------------------------
\subsection{Processing Cluster}
\label{sec:arch:cluster}

We combine a small 32\,bit RISC-V processor core (RV32IMC) \cite{Gautschi2017} with multiple NTX co-processors. Both operate on shared \SI{64}{\kilo\byte} \gls{tcdm} (reduced from \SI{128}{\kilo\byte} in \cite{schuiki2018scalable}). The memory is divided into 32 banks that are connected to the processors via an interconnect offering single-cycle access latency. An additional DMA engine allows the transfer of two-dimensional data planes between the \gls{tcdm} and the \gls{hmc}'s memory space. The RISC-V processors perform address calculation and control data movement via the DMA. Actual computation is performed on the data in the \gls{tcdm} by the NTX co-processors which we describe in the next section. An additional explicitly managed \SI{1.25}{\mega\byte} of memory outside the clusters holds the RISC-V binary executed by the processors and may be used by the program to cache frequently used data and shared variables.

% ------------------------------------------------------------------------------
\subsection{Network Training Accelerator (NTX)}
\label{sec:arch:nsl}

The computations involved in \glspl{dnn} training and many stencil codes are highly regular and can be broken down into a collection of reduction operations. The NTX co-processor is capable of performing thousands of \gls{fmac} cycles directly on the \gls{tcdm} without any RISC-V core intervention or explicit load or store instructions. The architecture of NTX is depicted in \figref{fig:arch:nsl}. It consists of four main blocks: the FPU containing the main data path, the register interface for command offloading, the controller that decodes the commands and issues micro-instructions to the FPU, and the address generators and hardware loops.

% ------------------------------------------------------------------------------
\subsection{FMAC and FPU}
\label{sec:arch:nsl:fmac}

The \gls{fpu} in NTX can perform fast \gls{fmac} operations with single-cycle throughput. It is based on a \gls{pcs} accumulator which aggregates the \SI{48}{\bit} multiplication result at full fixed-point precision ($\SI{\approx 300}{\bit}$). After accumulation the partial sums are reduced in multiple pipelined segments. The employed format has been aligned with IEEE\,754 32\,bit floats. The wide accumulator and deferred rounding allows NTX to achieve higher precision than conventional \glspl{fpu}. Analysis on a \gls{dnn} convolution layer has shown NTX's \gls{rmse} to be $1.7\times$ lower than that of a 32\,bit \gls{fpu}.

The \gls{fmac} unit allows NTX to handle common matrix operations such as inner/outer products and vector additions/multiplications. An additional comparator, index counter, and ALU register enable various additional commands such as finding minima/maxima, ReLU, thresholding and masking, and memcpy/memset \cite{schuiki2018scalable}.

% ------------------------------------------------------------------------------
\subsection{Hardware Loops and Address Generation}
\label{sec:arch:nsl:hwl}

\begin{figure}[t]
  \centering
  \includegraphics[width=\linewidth]{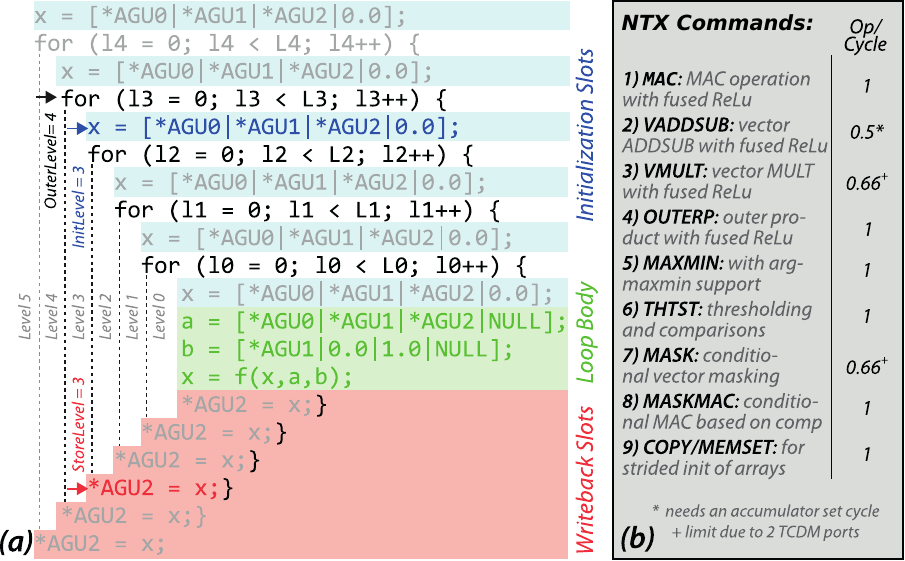}
  \caption{The structure of nested loops that can be directly offloaded to NTX (a), and an overview of the supported commands and their throughput (b). \texttt{[..|..]} indicates a choice of options; \texttt{*AGU0} indicates memory access at address \texttt{AGU0}.}
  \label{fig:arch:nsl:hwl}
\end{figure}

At the core of address generation in NTX are the five \glspl{hwl}. Each loop maintains a 16\,bit counter that has a programmable maximum count and can be enabled or disabled. The counters form a cascade to implement nested loops such that a loop wrapping from its maximum count to zero will increment the next higher loop. Three \glspl{agu} allow NTX to keep track of three pointers into memory. Each unit consists of a 32\,bit register holding the address and an adder. The address is incremented every cycle by one of five programmable step sizes chosen based on the outermost loop enabled in that cycle.

\figref{fig:arch:nsl:hwl}a shows the pseudo code structure of nested loops that NTX can natively perform. The number of loops (\emph{outer level}), position of the accumulator initialization (\emph{init level}), and position of the accumulator write back (\emph{store level}) are fully programmable. The \glspl{agu} provide addresses for the memory reads and writes depicted, thus removing the need for the majority of explicit load/store instructions. The operation performed by the FPU always occurs in the innermost loop and can be set to one of the commands listed in \figref{fig:arch:nsl:hwl}b.

% ------------------------------------------------------------------------------
\subsection{Offloading}
\label{sec:arch:nsl:offloading}

Each NTX has a set of configuration registers that are mapped into the memory space of the associated RISC-V core. As such the program can directly access and modify these registers, specifying the base address, strides, loop iterations, and command to be executed. Writing to the command register causes the current configuration to be copied into an internal buffer and executed, allowing the CPU to prepare the next command in parallel. All NTX attached to a core are aliased to a broadcast address, allowing efficient setting of common configuration values. This offloading scheme has proven to be very lean and efficient \cite{schuiki2018scalable}, allowing each NTX to run independently for thousands of cycles during which the RISC-V core can perform other tasks such as data movement.

We subdivide kernels to be executed into tiles. The DMA engine is used to copy input data into and results out of the \gls{tcdm} in a double buffering scheme, allowing the NTX co-processors to operate on one buffer while the DMA operates on another. This allows us to decouple and overlap computation and data movement, thus hiding memory latency and fully utilizing the available memory bandwidth.

%% file: sec_results.tex
% ==============================================================================
\section{Evaluation and Results}

% ------------------------------------------------------------------------------
\subsection{Silicon Results}
\label{sec:results:silicon}

We have implemented and taped out an NTX cluster in \textsc{Globalfoundries'} 22FDX, a 22\,nm FD-SOI technology. \tabref{tbl:ntx_22fdx} summarizes the figures of merit of our implementation. Post-layout timings were extracted from Cadence Innovus and used in a back-annotated gate-level simulation to obtain a trace of the cluster performing computation and DMA operation. This trace was then used in Innovus alongside the design to estimate the power consumption. The cluster consists of one RI5CY \cite{Gautschi2017} processor core and eight NTX coprocessors which operate on \SI{64}{\kilo\byte} of \gls{tcdm}. A \SI{2}{\kilo\byte} instruction cache with linear prefetching is located between the processor and the memory interface. The \gls{tcdm} and NTX operate at \SI{1.25}{\giga\hertz} in the worst case (\SI{0.72}{\volt}, \SI{125}{\celsius}/\SI{-40}{\celsius}, SSG), while the RISC-V processor and remaining cluster runs at half the speed, \SI{625}{\giga\hertz}. In this corner the cluster occupies \SI{0.51}{\milli\meter\squared} at 59\% placement density while achieving a compute performance of \SI{20}{\GFLOPs} and a memory bandwidth of \SI{5}{\giga\byte\per\second}. Assuming typical silicon (\SI{0.8}{\volt}, \SI{25}{\celsius}, TT) the cluster consumes \SI{186}{\milli\watt} of power while performing a $3\times 3$ convolution, which yields an energy efficiency of \SI{108}{\GFLOPsW} or conversely \SI{9.3}{\pico\joule\per\FLOP}.

\input{tables/ntx_22fdx}

\begin{figure}
  \centering
  \includegraphics[width=\linewidth]{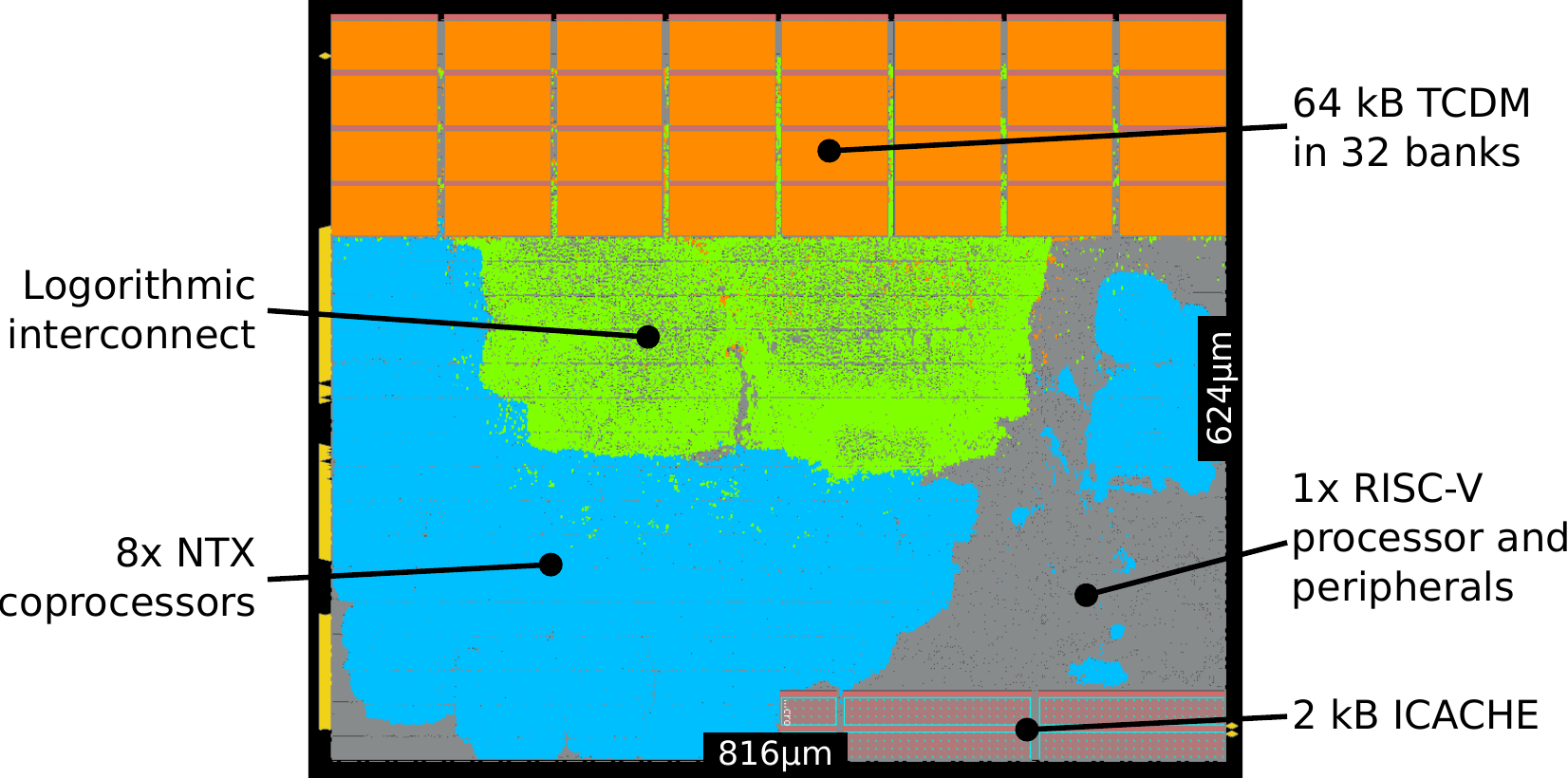}
  \caption{Floorplan of the NTX cluster as implemented in 22FDX. Highlighted are the main area contributors: TCDM, logarithmic interconnect, RISC-V processor and peripherals, NTX coprocessors, and the instruction cache.}
  \label{fig:floorplan}
\end{figure}

% ------------------------------------------------------------------------------
\subsection{Evaluated Kernels}
\label{sec:results:kernels}

We estimate the execution time of a kernel based on the model presented in \cite{schuiki2018scalable}. The data is assumed to initially reside outside the cluster, e.g. in a DRAM attached to the AXI port.

\subsubsection{Basic Linear Algebra Subprograms}

The AXPY ($y = a\cdot x + y$), matrix-vector product GEMV, and matrix-matrix product GEMM are taken from the BLAS 1, 2, and 3 set of kernels, respectively. For AXPY and GEMV the input data is split into tiles that fit into the cluster's TCDM memory, which are then processed tile-by-tile. Data reuse, which manifests itself as increased operational intensity, is limited by the kernel itself as well as the size of the largest tile that fits into the TCDM. For GEMM we use a block matrix multiplication to subdivide the input matrices.

\subsubsection{Convolutions}

We evaluate $3\times 3$, $5\times 5$, and $7\times 7$ convolutions as they commonly appear in \glspl{dnn} \cite{Szegedy2015}. Reuse factors per image pixel are 9, 25, and 49, respectively. Larger convolution kernels exhibit higher operational intensity since input image pixels are reused for more operations, thus allowing NTX to operate even further in the compute-bound regime.

\subsubsection{Stencils}

Stencil codes are common in \gls{hpc}. We evaluate the Discrete Laplace Operator \cite{wardetzky2007discrete} in one, two, and three dimensions with three, five, and seven coefficients, respectively. Its star shaped access pattern allows it to be computed efficiently on NTX by decomposing the kernel into its separate dimensions. Furthermore, we also consider the diffusion kernel presented as an example in \cite{gysi2015modesto} which has 13 coefficients and can be decomposed into three separate NTX instructions with nine, two, and two coefficients each. Together with convolutions these are a representative sample of the common five- and nine-point stencil shapes and beyond.

% ------------------------------------------------------------------------------
\subsection{Roofline Model}
\label{sec:results:roofline}

\begin{figure}
  \centering
  \includegraphics[width=\linewidth]{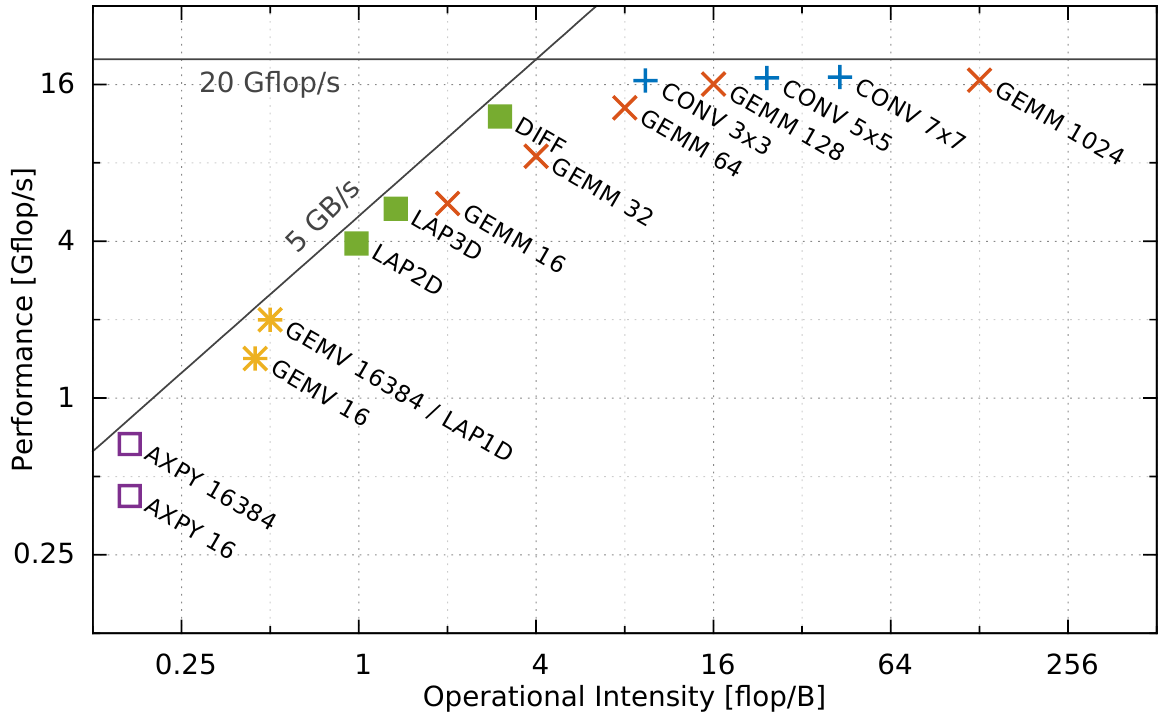}
  \caption{Roofline model of NTX for different kernels. For AXPY, GEMV, and GEMM the vector length and matrix side length are indicated. For CONV the size of the convolution kernel is indicated. LAP are discrete Laplace operators in 1D, 2D, and 3D. DIFF is the diffusion stencil used as an example in \cite{gysi2015modesto}. Note that LAP1D coincides with GEMV 16384.}
  \label{fig:roofline}
\end{figure}

The roofline model of one NTX cluster is depicted in \figref{fig:roofline}. The eight NTX co-processors at \SI{1.25}{\giga\hertz} achieve \SI{20}{\GFLOPs} of peak performance, while the AXI memory port at \SI{625}{\mega\hertz} can carry \SI{5}{\giga\byte\per\second} of peak traffic. We estimate the performance of different kernels by extrapolation of a gate-level simulation of the $3\times 3$ convolution. For the three BLAS kernels AXPY, GEMV, and GEMM the NTX cluster achieves close to maximum performance with a sufficiently large problem size. AXPY and GEMV are memory bound in all configurations, while GEMM quickly becomes compute bound as operational intensity increases due to better amortization of constant setup and write back overheads. The investigated convolution kernels are all compute bound and achieve close to maximum performance. The three Discrete Laplace Operator \cite{wardetzky2007discrete} and the diffusion stencil \cite{gysi2015modesto} are all memory bound, yet achieve close to maximum bandwidth utilization since their regular structure is highly amenable to execution on NTX.

We observe in simulations that the practically achievable compute performance is limited by the probability of a banking conflict in the TCDM interconnect, which causes an NTX stall. This probability was measured to be around 13\%, which puts the maximum performance achievable in practice at around \SI{17.4}{\GFLOPs}. This also limits the expected maximum memory bandwidth of the system for memory-bound kernels to around \SI{4.35}{\GFLOPs}.

The memory bound of the roofline plot is dictated by the width of the AXI port of the cluster, which is a design parameter. It was set to 64\,bit to accommodate the bandwidth requirements of \gls{dnn} training and to facilitate system integration also with lower-end devices. This parameter could be increased to 128 or 256\,bit, raising the bandwidth limit to \SI{10}{\giga\byte\per\second} and \SI{20}{\giga\byte\per\second}, respectively. This would allow the cluster to sustain very high utilization also for operational intensities down to \SI{2}{\FLOP\per\byte} and \SI{1}{\FLOP\per\byte}, respectively.

% ------------------------------------------------------------------------------
\subsection{Neural Network Training Efficiency}
\label{sec:results:comp_others}

\input{tables/system_comp}

\begin{figure}
  \centering
  \includegraphics[width=\linewidth]{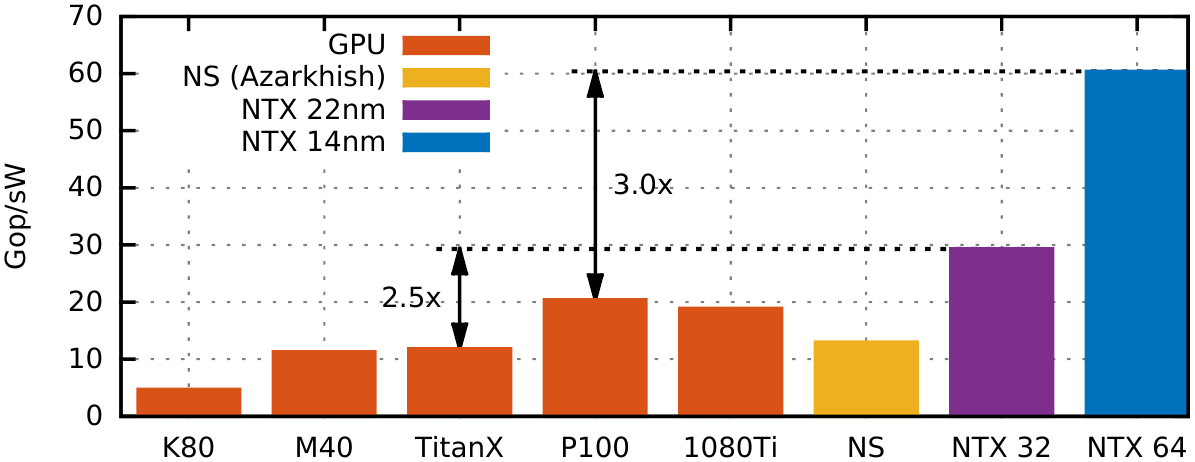}
  \caption{Comparison of energy efficiency when training the networks listed in \tabref{tbl:results:syscmp} (geometric mean), with GPUs, NS \cite{azarkhish2017neurostream}, and the largest NTX configurations that do not require additional \glspl{lim}. NTX\,32 in 22\,nm achieves a $2.5\times$ increase, and NTX\,64 in 14\,nm a $3\times$ increase in efficiency over GPUs in similar technology nodes.}
  \label{fig:results:eff_vs_net}
\end{figure}

To compare \gls{dnn} training performance against other accelerators we reproduce \tabref{tbl:results:syscmp} and \figref{fig:results:eff_vs_net} from \cite{schuiki2018scalable} with updated numbers based on our implementation in 22FDX, which provides a more accurate estimate of the performance achievable with the resulting hardware. Among the custom accelerators, DaDianNao has an efficiency of \SI{65.8}{\GOPsW} with fixed-point arithmetic, which is similar to the computationally equivalent NTX~128. ScaleDeep has an efficiency of \SI{100.8}{\GFLOPsW} which is $1.3\times$ higher than NTX~512, the largest configuration considered by us. Furthermore our architecture can achieve significantly higher energy efficiency than a GPU at a comparable technology node (see \figref{fig:results:eff_vs_net}). Considering the largest NTX configurations that do not require additional \glspl{lim}, we achieve an efficiency increase of $2.5\times$ from \SI{11.8}{\GOPsW} to \SI{29.5}{\GOPsW} in \SI{22}{\nano\meter}, and an increase of $3\times$ from \SI{20.4}{\GOPsW} to \SI{63.5}{\GOPsW} in \SI{14}{\nano\meter}.

\begin{figure}
  \centering
  \includegraphics[width=\linewidth]{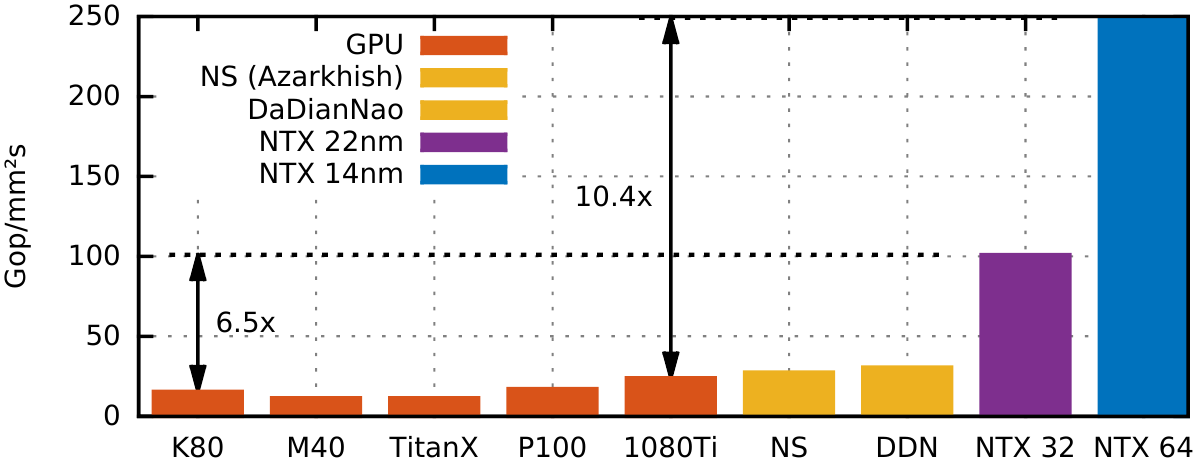}
  \revb{\caption{Comparison of the \si{\GOPs} of compute performance per deployed area of silicon, for GPUs, NS \cite{azarkhish2017neurostream}, and the largest NTX configurations that do not require additional \glspl{lim}. NTX\,32 in 22\,nm achieves a $6.5\times$ increase, and NTX\,64 in 14\,nm a $10.4\times$ increase in area efficiency over GPUs in similar technology nodes.}}
  \label{fig:results:deployed_si}
\end{figure}

We also reproduce \figref{fig:results:deployed_si} from \cite{schuiki2018scalable} with updated numbers for 22FDX to compare the \si{\GOPs} of compute performance per deployed amount of silicon for NTX and GPUs. Our solution requires $10.4\times$ less area to achieve the same compute performance as a GPU.

%% file: tables/ntx_22fdx.tex
\begin{table}
\begin{threeparttable}
\caption{Figures of merit of one NTX cluster implemented in 22FDX.}
\label{tbl:ntx_22fdx}
\begin{tabularx}{\linewidth}{@{}XlXl@{}}
\toprule
    \textbf{Processors:} & 1 RISC-V & \textbf{Memory:} & \SI{64}{\kilo\byte} TCDM \\
                         & 8 NTX    &                  & \SI{2}{\kilo\byte} ICACHE \\

    \textbf{Frequency:} & \SI{1.25}{\giga\hertz} NTX    & \textbf{Area:} & \SI{0.51}{\milli\meter\squared} \\
                        & \SI{625}{\mega\hertz} Cluster &                & 59\% density\\

    \textbf{Peak Perf.:} & \SI{20}{\GFLOPs}              & \textbf{Power:}      & \SI{186}{\milli\watt} \\
                         & \SI{5}{\giga\byte\per\second} & \textbf{Efficiency:} & \SI{108}{\GFLOPsW} \\
                         &                               &                      & \SI{9.3}{\pico\joule\per\FLOP} \\
\bottomrule
\end{tabularx}
\end{threeparttable}
\end{table}

%% file: tables/system_comp.tex
\begin{table*}
\newcommand\nbold{}
\begin{threeparttable}
\caption{Comparison between different configurations of the architecture proposed in this work, related custom accelerators, and GPUs. The energy efficiencies reported are with respect to training different \glspl{dnn}. Refer to \cite{schuiki2018scalable} for a more detailed treatment.}
\label{tbl:results:syscmp}
\renewcommand*{\arraystretch}{0.9}
% \small
\begin{tabularx}{\linewidth}{@{}X llk{3.1}ck{1.2}k{3.3}c *{7}{k{3.1}}@{}}
\toprule

  \textbf{Platform} &
  \multicolumn{7}{c}{\textbf{Characteristics}} &
  \multicolumn{7}{c@{}}{\textbf{Energy Efficiency} [\si{\GOPsW}]} \\

  \cmidrule(lr){2-8}
  \cmidrule(l){9-15}

  &
  \rotatebox{90}{Logic [\si{\nano\metre}]} &
  \rotatebox{90}{DRAM [\si{\nano\metre}]} &
  \rotatebox{90}{Area [\si{\milli\meter\squared}]} &
  \rotatebox{90}{\gls{lim}} &
  \rotatebox{90}{Freq. [\si{\giga\hertz}]} &
  \rotatebox{90}{Peak \si{\tera\op\per\second}} &
  \rotatebox{90}{Arithmetic} &
  \rotatebox{90}{AlexNet \cite{Krizhevsky2012}} &
  \rotatebox{90}{GoogLeNet \cite{Szegedy2015}} &
  \rotatebox{90}{Incep. v3 \cite{Szegedy2016}} &
  \rotatebox{90}{ResNet 34 \cite{he2016deep}} &
  \rotatebox{90}{ResNet 50 \cite{he2016deep}} &
  \rotatebox{90}{ResNet 152 \cite{he2016deep}} &
  \rotatebox{90}{\textbf{Geom. Mean}} \\

\midrule

  \multicolumn{7}{@{}l}{\textbf{This Work}} \\
  NTX (16$\times$)  & 22 & 50 & \ubold  4.8 & 0 & 2.50 & 0.640 & (a) &  19.8 &  23.7 &  24.3 &  21.7 &  21.4 &  23.6 & \ubold 22.5 \\
  NTX (32$\times$)  & 22 & 50 & \ubold  9.6 & 0 & 1.90 & 0.973 & (a) &  25.8 &  30.9 &  31.6 &  28.2 &  27.9 &  30.8 & \ubold 29.3 \\
  NTX (64$\times$)  & 22 & 50 & \ubold 19.3 & 1 & 1.43 & 1.466 & (a) &  32.3 &  38.8 &  39.7 &  35.4 &  35.0 &  38.6 & \ubold 36.7 \\
  NTX (16$\times$)  & 14 & 30 & \ubold  1.9 & 0 & 3.50 & 0.896 & (a) &  31.6 &  37.9 &  38.8 &  34.6 &  34.2 &  37.7 & \ubold 35.9 \\
  NTX (32$\times$)  & 14 & 30 & \ubold  3.9 & 0 & 2.66 & 1.362 & (a) &  41.8 &  50.1 &  51.3 &  45.8 &  45.2 &  49.9 & \ubold 47.5 \\
  NTX (64$\times$)  & 14 & 30 & \ubold  7.7 & 0 & 1.88 & 1.920 & (a) &  53.2 &  63.8 &  65.3 &  58.3 &  57.6 &  63.5 & \ubold 60.4 \\
  NTX (128$\times$) & 14 & 30 & \ubold 15.4 & 1 & 0.94 & 1.920 & (a) &  62.1 &  74.6 &  76.2 &  68.1 &  67.2 &  74.2 & \ubold 70.6 \\
  NTX (256$\times$) & 14 & 30 & \ubold 30.8 & 2 & 0.47 & 1.920 & (a) &  66.9 &  80.3 &  82.1 &  73.3 &  72.4 &  79.8 & \ubold 76.0 \\
  NTX (512$\times$) & 14 & 30 & \ubold 61.6 & 3 & 0.23 & 1.920 & (a) &  69.3 &  83.2 &  85.0 &  75.9 &  75.0 &  82.7 & \ubold 78.7 \\

\midrule

  \multicolumn{7}{@{}l}{\textbf{Custom Accelerators}} \\
  NS (16$\times$) \cite{azarkhish2017neurostream} & 28 & 50    & \ubold  9.3 & {---} & 1.0 & 0.256 & (a) & 10.2            & 15.1            & 14.6  & 13.1             & 12.9  & 14.2  & \ubold 13.0 \\
  DaDianNao \cite{luo2017dadiannao}               & 28 & 28    & \ubold 67.7 & {---} & 0.6 & 2.09  & (b) & {---}           & {---}           & {---} & {---}            & {---} & {---} & \ubold 65.8 \\
  ScaleDeep \cite{venkataramani2017scaledeep}     & 14 & {---} & {---}       & {---} & 0.6 & 680   & (c) & 87.7 & 83.0 & {---} & 139.2 & {---} & {---} & \ubold 100.8 \\

\midrule

  \multicolumn{7}{@{}l}{\textbf{GPUs}} \\
  Tesla K80    & 28 & 40 & \ubold 561 & {---} & 0.59 & 8.74 & (a) & {---} & 4.5  & 3.5   & {---} & 3.7   & 8.8   & \ubold  4.7 \\
  Tesla M40    & 28 & 30 & \ubold 601 & {---} & 1.11 & 7.00 & (a) & {---} & 11.3 & {---} & {---} & {---} & {---} & \ubold 11.3 \\
  Titan X      & 28 & 30 & \ubold 601 & {---} & 1.08 & 7.00 & (a) & 12.8  & 9.9  & {---} & 17.6  & 8.5   & 12.2  & \ubold 11.8 \\
  Tesla P100   & 16 & 21 & \ubold 610 & {---} & 1.3  & 10.6 & (a) & {---} & 19.8 & 19.5  & {---} & 18.6  & 24.18 & \ubold 20.4 \\
  GTX 1080 Ti  & 16 & 20 & \ubold 471 & {---} & 1.58 & 11.3 & (a) & 20.1  & 16.6 & {---} & 27.6  & 13.4  & 19.56 & \ubold 18.9 \\

\bottomrule
\end{tabularx}
\end{threeparttable}
\end{table*}

%% file: sec_relwork.tex
% ==============================================================================
\section{Related Work}
\label{sec:relwork}

Acceleration of \glspl{dnn} is a well researched field with a rich literature. Goodfellow, et al.\@ \cite{Goodfellow2016} provide a good coverage of the mathematical background of Deep Learning, and \cite{Sze2017} offer an overview of techniques for efficient \gls{dnn} inference and the challenges involved. Dedicated \gls{dnn} accelerators have mainly focused on inference \cite{Du2015, Gao2017, azarkhish2017neurostream}. The increasing size of parameter and training data of state-of-the-art networks \cite{Szegedy2015, he2016deep} provide a compelling reason for \gls{pim} solutions. We observe that fewer architectures that support both inference and training have been proposed so far \cite{kim2016neurocube, luo2017dadiannao, venkataramani2017scaledeep}. DaDianNao \cite{luo2017dadiannao} is a multi-node system achieving an energy efficiency of around \SI{350}{\GOPsW} for \SI{16}{\bit} fixed point arithmetic. ScaleDeep \cite{venkataramani2017scaledeep} is a multi-node system with heterogeneous chips assembled from memory-heavy and compute-heavy tiles that distribute the \gls{dnn} state across several chips and nodes, achieving a very high energy efficiency around \SI{332}{\GFLOPsW} in a \SI{14}{\nano\metre} technology. A more detailed treatment of related accelerators can be found in \cite{schuiki2018scalable}.

GPUs are commonly used for both inference and training, where recent implementations on the GTX~780 and GTX~Titan reach \SI{1650}{\GFLOPs} at \SI{250}{\watt} and \SI{999}{\GFLOPs} at \SI{240}{\watt}, which corresponds to 6.6 and \SI{4.2}{\GFLOPsW}, respectively \cite{Cavigelli2015b, azarkhish2017neurostream}. Embedded GPUs like the Tegra K1 have lower absolute throughput, but reach a energy efficiencies of around \SI{7}{\GFLOPsW} \cite{Cavigelli2015b}. Newer GPU generations such as Pascal offer \gls{hbm} and 16\,bit \gls{fp} support allowing for higher peak throughput and efficiency of up to \SI{10.6}{\TFLOPs} and \SI{20}{\GFLOPsW}, respectively \cite{johnson2017, TfBench2017}.

% The recently introduced Volta generation offers \emph{tensor cores}, a new compute element able to perform 4$\times$4 matrix \glspl{fma} in 16\,bit \gls{fp}, with 16 or 32\,bit outputs in one cycle. These cores promise a $5\times$ increase in Deep Learning performance compared to previous GPU generations \cite{nvidia2017volta}.

In the \gls{hpc} domain stencil codes and general linear algebra are crucial building blocks of many applications. The increasing complexity and data volume of state-of-the-art problems requires dedicated acceleration engines to keep power consumption manageable. Green Wave \cite{krueger2011hardware} for example focuses on solving 8th order Laplacian stencils for seismic modeling applications using a large array of dedicated streaming nodes, reaching \SI{82.5}{\GFLOPs} at \SI{1.25}{\GFLOPsW}. A GPU executing the same stencil reaches \SI{145}{\GFLOPs} at \SI{0.33}{\GFLOPsW}, a $1.7\times$ increased performance but at the cost of $3.5\times$ less energy efficiency \cite{krueger2011hardware}. We estimate NTX\,16 to achieve \SI{130}{\GFLOPs} at \SI{11}{\GFLOPsW} on the same stencil code. This suggests that dedicated streaming-based accelerators for stencil codes and linear algebra are an attractive proposition to reduce the energy footprint of \gls{hpc} applications.

%% file: sec_conc.tex
\section{Conclusion}
\label{sec:conc}

We have presented an evaluation of the NTX floating point co-processor \cite{schuiki2018scalable} based on a concrete implementation taped out in \textsc{Globalfoundries'} 22FDX technology. Power analysis based on post-layout simulation confirms the estimates in previous work. The hardware loop and \gls{fmac} capabilities of NTX apply well to kernels beyond \glspl{dnn} such as stencil codes prevalent in \gls{hpc}, while allowing NTX to achieve very high utilization of its peak performance, and suggests that the co-processor is capable of handling other well-structured problems to be investigated further.